\documentclass[camera ready]{Interspeech}
\usepackage{amsmath}

\title{Bridging Self-Supervised Learning and Speech Enhancement: A Wav2Vec2-Conditioned Framework}

\author[affiliation={1}, orcid=0009-0009-3199-104X, correspondingauthor]{Shuubham}{Ojha}
\author[affiliation={1}, orcid=0000-0002-1012-183X, ]{Carol}{Espy-Wilson}
\address{
    $^1$ Institute for Systems Research, University of Maryland, College Park 
}

\email{sojha1@umd.edu, espy@umd.edu}


\usepackage{comment}
\usepackage{graphicx}
\usepackage{subcaption}

\begin{document}

\maketitle

\begin{abstract}
Diffusion models show potential for speech enhancement but lack linguistic guidance. We condition a diffusion-based model on wav2vec 2.0 features from noisy input, injected at the U-Net bottleneck via Feature-wise Linear Modulation (FiLM). Phonetic representations from wav2vec 2.0 features of degraded speech, anchor the reverse diffusion process. While a frozen wav2vec 2.0 encoder extracts features, a learned FiLM generator produces scale and shift parameters modulating the bottleneck with minimal overhead. Motivated by the optimal Bayesian causal estimator under a linear-Gaussian state-space model, FiLM coefficients are aggregated via exponential smoothing for temporal compression. Evaluation on VoiceBank-DEMAND and LibriMix shows competitive performance against the unconditioned baseline in PESQ, STOI, SI-SDR and DNSMOS. We consistently record an improvement of 0.4 on PESQ score, suggesting self-supervised representations effectively condition diffusion-based speech enhancement.
 
\end{abstract}

\section{Introduction}
Speech enhancement (SE) seeks to recover clean speech from signals degraded by additive noise, reverberation, and other distortions~\cite{loizou2013}. To this end, discriminative methods that learn a direct mapping from noisy to clean speech, whether through time-frequency masking~\cite{wang2014}, complex spectral mapping~\cite{williamson2016}, or time-domain regression~\cite{luo2019}, have been the dominant approach. Yet these methods often yield over-smoothed outputs and can introduce audible artifacts, particularly at low signal-to-noise ratios (SNRs).
Diffusion-based generative models offer a promising alternative by learning to reverse a gradual noising process, producing enhanced speech that sounds more natural~\cite{ho2020,song2021sde}. CDiffuSE~\cite{lu2022cdiffuse} was among the first to apply denoising diffusion probabilistic models (DDPMs) to SE. Subsequent works have built on this idea: SGMSE and SGMSE+~\cite{welker2022,richter2023} operate in the complex STFT domain using interpolated forward processes, while UNIVERSE~\cite{serra2022} employs a separate conditioning network to handle a wide range of distortion types. Although these generative methods achieve high perceptual quality when training and test conditions are well matched, they often struggle to generalize to unseen noise types or acoustic settings~\cite{richter2023}.
A key design decision in diffusion-based SE is how to supply conditioning information to steer the reverse process. Common strategies include input concatenation~\cite{lu2022cdiffuse} and dedicated conditioning networks~\cite{serra2022,scheibler2024}. On a different note, self-supervised learning (SSL) models such as wav2vec 2.0~\cite{baevski2020}, HuBERT~\cite{hsu2021}, and WavLM~\cite{chen2022wavlm} have been found to capture rich phonetic and linguistic features that remains informative even under noisy conditions. Several studies have explored using SSL representations for SE via auxiliary losses~\cite{huang2022,hung2022}, as input features~\cite{chen2023,andreev2024finally}, or as discrete tokens for language-model-based enhancement~\cite{yang2023}. 
In this work, we propose conditioning a diffusion-based SE model on wav2vec 2.0 features extracted from the noisy input, injected at the U-Net bottleneck through Feature-wise Linear Modulation (FiLM)~\cite{perez2018}. To produce a temporally compressed conditioning signal within the memory constraints of diffusion-based training, we apply exponential averaging to the projected FiLM coefficients. Our main contributions are: 
\begin{itemize}
    \item a FiLM-based conditioning mechanism that incorporates wav2vec 2.0 features at the U-Net bottleneck for diffusion-based SE,
    \item a principled temporal smoothing strategy for the conditioning signal, and, evaluation on standard benchmarks with improvements on PESQ, STOI and DNSMOS with marginal tradeoffs in SI-SDR.
    % \item evaluation on standard benchmarks with improvements on PESQ, STOI and DNSMOS with marginal tradeoffs in SI-SDR. 
    % \item a systematic evaluation on the VoiceBank-DEMAND and LibriMix benchmarks showing improvements across all metrics for LibriMix and consistent gains in PESQ and STOI over unconditioned baselines with marginal trade-offs in SI-SDR and DNSMOS OVRL for VoiceBank-DEMAND for a lightweight version of our model.
\end{itemize}

% This operation suppresses frame-level noise in the wav2vec 2.0 features without introducing future context, making it compatible with causal inference.

% To produce a temporally compressed conditioning signal within the memory limits of diffusion-based training, we apply exponential smoothing to the projected FiLM coefficients, a choice we justify as the optimal Bayesian causal estimator under a linear-Gaussian state-space model of the underlying phonetic dynamics~\cite{meinhold1983,harvey1990}. 

\section{Related Work}

\subsection{Conditional Diffusion Models}

Diffusion models~\cite{ho2020,song2021sde} generate data by reversing a gradual noising process. To steer generation toward a desired output, conditioning can be introduced via classifier guidance~\cite{dhariwal2021}, classifier-free guidance~\cite{ho2022cfg}, input concatenation, cross-attention~\cite{rombach2022}, or adaptive normalization~\cite{peebles2023,dhariwal2021}. Feature-wise Linear Modulation (FiLM)~\cite{perez2018} formalizes the last approach, applying a learned affine transformation $\gamma \odot \mathbf{h} + \beta$ to feature maps $\mathbf{h}$, where $\gamma$ and $\beta$ are predicted from the conditioning signal. FiLM and its variants (e.g., AdaLN-Zero~\cite{peebles2023}) have been widely adopted in diffusion architectures for vision.

\subsection{Diffusion-Based Speech Enhancement}

StoRM~\cite{lemercier2023storm} adopts a score-based diffusion framework in the complex STFT domain, using an Ornstein-Uhlenbeck Variance Exploding (OUVE) process. The forward process interpolates between the clean STFT $\mathbf{x}_0$ and the noisy observation $\mathbf{y}$ via a perturbation kernel:
$$p_{0,t}(\mathbf{x}_t \mid \mathbf{x}_0, \mathbf{y}) = \mathcal{N}_{\mathbb{C}}\left(\mathbf{x}_t;\; e^{-\gamma t}\mathbf{x}_0 + (1 - e^{-\gamma t})\mathbf{y},\; \sigma(t)^2\mathbf{I}\right)$$
where $\gamma$ controls the decay rate toward $\mathbf{y}$ and $\sigma(t)^2$ is a monotonically increasing variance schedule. A score network $s_\theta(\mathbf{x}_t, \mathbf{y}, t)$ is trained via denoising score matching and a reverse-time stochastic differential equation (SDE), utilizing this trained score function is solved at inference to produce the enhanced waveform. We build directly on this framework.

\subsection{Self-Supervised Representations for Speech Enhancement}

Self-supervised models such as wav2vec~2.0~\cite{baevski2020}, HuBERT~\cite{hsu2021}, and WavLM~\cite{chen2022wavlm} learn rich phonetic representations that remain informative even under noisy conditions. Prior work has leveraged SSL features for SE via auxiliary losses~\cite{hung2022}, as input features \cite{chen2023,andreev2024finally},  or FiLM with discriminative latents in a GAN framework~\cite{discogan2025}. Our work differs in injecting temporally-smoothed wav2vec~2.0 embeddings into a diffusion backbone via FiLM, with a theoretically motivated smoothing strategy.

\begin{table}[t]
\centering
\caption{Intrusive metrics on VB-DEMAND test set. Best in \textbf{bold}, second-best \underline{underlined}. Each variant of StoRM and the proposed model (except Mean Pool) are trained for 100 epochs. Mean Pool achieved convergence after 81 epochs.}
\label{tab:seen_intrusive}
\begin{tabular}{lccc}
\toprule
\textbf{Model} & \textbf{PESQ\,$\uparrow$} & \textbf{STOI\,$\uparrow$} & \textbf{SI-SDR\,$\uparrow$} \\
\midrule
Noisy           & 1.9797 & 0.7867 & 8.4450 \\
CDiffuSE (large)~\cite{lu2022cdiffuse}  & 2.52 & - & - \\
% NASE~\cite{hu2024nase}          & \textbf{2.98} & \textbf{0.88} & \textbf{18.9} \\
UNIVERSE \cite{serra2022} & 2.55 & 0.784 &  -\\ 
UNIVERSE++ \cite{scheibler2024} & 2.88 & \underline{0.860} & - \\
StoRM-128~\cite{lemercier2023storm} & 2.4862 & 0.8571 & 1\textbf{8.5656} \\
StoRM-32~\cite{lemercier2023storm}  & 2.7941 & 0.8522 & 17.9520 \\
% FINALLY~\cite{andreev2024finally}   & x.xx & x.xx & x.xx \\
\midrule
\textsc{Ours-128} & 2.8742 & \textbf{0.8673} & \underline{17.9844} \\
\textsc{Ours-32}  & \underline{2.8636} & 0.8589 & 17.8179 \\
\textsc{Mean Pool-32 (OURS)} & \textbf{2.9186} & 0.8613 & 17.3581\\
\bottomrule
\end{tabular}
\end{table}

\section{Proposed Architecture}

\begin{table}[t]
\centering
\caption{Non-intrusive DNSMOS metrics on VB-DEMAND test set.}
\label{tab:seen_dnsmos}
\begin{tabular}{lccc}
\toprule
\textbf{Model} & \textbf{SIG\,$\uparrow$} & \textbf{BAK\,$\uparrow$} & \textbf{OVRL\,$\uparrow$} \\
\midrule
Noisy           & 3.249  & 2.878 & 2.588 \\
CDiffuSE (large)~\cite{lu2022cdiffuse}  & \textbf{3.72} & 2.91 & 3.10 \\
% NASE~\cite{hu2024nase}          & - & - & - \\
UNIVERSE \cite{serra2022} & - & - & 3.12 \\ 
UNIVERSE++ \cite{scheibler2024} & - & - & 3.20 \\
StoRM-128~\cite{lemercier2023storm} & 3.624 & 3.724 & 3.229 \\
StoRM-32~\cite{lemercier2023storm}  & 3.606 & \textbf{4.073} & \underline{3.347}\\
% FINALLY~\cite{andreev2024finally}   & x.xx & x.xx & x.xx \\
\midrule
\textsc{Ours-128} & 3.608 & 3.952 & 3.300 \\
\textsc{Ours-32}  & \underline{3.636} & \underline{3.968} & \textbf{3.359} \\
\textsc{Mean Pool-32 (OURS)} & 3.557 & 3.939 & 3.253\\
\bottomrule
\end{tabular}
\end{table}

\subsection{U-Net Score Network}

The score network follows the NCSN++ architecture~\cite{song2021ncsn}, a U-Net with BigGAN-style residual blocks, skip connections, and self-attention layers at selected resolutions. The encoder progressively downsamples the noisy STFT representation through a series of residual blocks, producing feature maps of decreasing spatial resolution and increasing channel dimension. The bottleneck processes the most compressed representation, and the decoder mirrors the encoder with upsampling and skip connections from corresponding encoder levels.
The diffusion timestep $t$ is embedded via a sinusoidal positional encoding followed by a learned linear projection, and injected into each residual block through additive bias after the first convolution. The noisy speech STFT $\mathbf{y}$ is provided as an additional input channel to the U-Net.

\begin{table}[t]
\centering
\caption{Ablation on smoothing coefficient $\alpha$ for PESQ scores (VB-DEMAND, 128-ch).}
\label{tab:ablation_alpha}
\begin{tabular}{lcccc}
\toprule
$\alpha$ & \textbf{2.5dB} & \textbf{7.5dB} & \textbf{12.5dB} & \textbf{17.5dB} \\
\midrule
0.0   & 2.4455 & \textbf{2.8393} & 3.0019 & 3.2490\\
0.01  & 2.4334 & 2.8252 & 3.0133 & 3.2388\\
0.50  & 2.4394 & 2.8330 & 3.0194 & 3.2636\\
0.75  & 2.4434 & 2.8285 & 2.9968 & 3.2589\\
1.00  & \textbf{2.4580} & 2.8225 & \textbf{3.0304} & \textbf{3.2855}\\
%Mean pooling          & x.xx & x.xx \\
\bottomrule
\end{tabular}
\end{table}

\subsection{Wav2Vec 2.0 Feature Extraction}

We extract wav2vec 2.0 representations from the noisy waveform $y(n)$ using a pretrained and frozen wav2vec 2.0 base model~\cite{baevski2020}. The model consists of a multi-layer convolutional feature encoder that produces latent speech representations at a 20ms frame rate, followed by a Transformer encoder that contextualizes these representations over the full utterance.

We use the output of the final Transformer layer as our conditioning features, yielding a sequence $\mathbf{W} = [\mathbf{w}_1, \ldots, \mathbf{w}_L] \in \mathbb{R}^{L \times D}$, where $L$ is the number of frames and $D = 768$ is the feature dimension. Although these features are extracted from noisy speech, prior work has shown that wav2vec 2.0 representations retain substantial phonetic and linguistic information even under degraded conditions~\cite{baevski2020,hsu2021robust}, owing to the model's pretraining on diverse unlabeled speech data with a contrastive objective that encourages invariance to surface-level acoustic variation.

\begin{table}[t]
\centering
\caption{Inference time comparison. RTF $< 1$ indicates faster-than-real-time processing. T. Params denotes number of trainable parameters.}
\label{tab:inference}
\begin{tabular}{lccc}
\toprule
\textbf{Model} & \textbf{T. Params (M)} & \textbf{RTF\,$\downarrow$} & \textbf{Real Time} \\
\midrule
StoRM-128~\cite{lemercier2023storm} & 55.1 & 1.63 & No \\
\textsc{Ours-128}                    & 55.1 & 1.8 & No \\
\midrule
StoRM-32~\cite{lemercier2023storm}  & 3.6 & \textbf{0.36} & Yes \\
\textsc{Ours-32}                     & 3.6 & \textbf{0.55} & Yes \\
\bottomrule
\end{tabular}
\end{table}

\subsection{FiLM Conditioning at the Bottleneck}

We condition the U-Net on the wav2vec 2.0 features using Feature-wise Linear Modulation (FiLM)~\cite{perez2018} applied at the bottleneck. The wav2vec 2.0 features are projected to the bottleneck feature space via 3    cascaded fully connected layers to reduce it's size from  $\mathbb{R}^{T' \times 768}$  to $\mathbb{R} ^ {2 \times T' \times C}$, where $C = 64$ or 256 (for 32 and 128-ch configurations resp (see Section \ref{sec:Dataset_and_setup})). In this compressed representation, the two channels are identified with $\gamma$ and $\beta$ respectively. Upon temporal smoothing (see Sec. \ref{sec:temporal_smoothing}), $\gamma$ and $\beta$ are compressed to $\tilde{\gamma}$ and $\tilde{\beta}$ respectively each of size $\mathbb{R}^{1 \times C}$. Given bottleneck features $\mathbf{h} \in \mathbb{R}^{H \times W \times C}$ and the projected wav2vec 2.0 features, the modulated bottleneck features are then computed as:

$$\tilde{\mathbf{h}} = \tilde{\gamma} \odot \mathbf{h} + \tilde{\beta}$$

where $\odot$ denotes element-wise multiplication along the channel dimension. We choose to apply FiLM at the bottleneck rather than at all U-Net levels for two reasons. First, the bottleneck contains the most abstract and compressed representation, making it a natural point to inject high-level semantic information from wav2vec 2.0. Second, modulating only the bottleneck keeps the computational overhead minimal. We prove through an ablation study that over conditioning leads to inferior enhancement performance.

\subsection{Temporal Aggregation of FiLM Coefficients}
\label{sec:temporal_smoothing}
The FiLM coefficients produced by the linear projection of wav2vec 2.0 features form a sequence over time. To condition the bottleneck with a single scale-shift pair, we require a principled strategy for aggregating this sequence. We derive the functional form of our aggregation by considering the optimal causal estimator under a linear-Gaussian state-space model.

Consider a scalar model for a single FiLM coefficient (the argument applies identically to every element of $\gamma_n$ and $\beta_n$). Let $c_n$ denote the underlying phonetic conditioning state at frame $n$, and let $\tilde{c}_n$ denote the projected FiLM coefficient at that frame. We posit the following state-space model:
\begin{equation}
    c_n = c_{n-1} + w_n, \quad w_n \sim \mathcal{N}(0, \sigma_w^2), \label{eq:state}
\end{equation}
\begin{equation}
    \tilde{c}_n = c_n + v_n, \quad v_n \sim \mathcal{N}(0, \sigma_v^2), \label{eq:obs}
\end{equation}
where the state equation~\eqref{eq:state} models the phonetic content as a random walk with process noise variance $\sigma_w^2$, and the observation equation~\eqref{eq:obs} treats each projected coefficient $\tilde{c}_n$ as a noisy measurement of $c_n$ with observation noise variance $\sigma_v^2$. The noise sequences $w_n$ and $v_n$ are assumed mutually independent.

% Since both equations are linear with additive Gaussian noise, 
The MMSE causal estimator of $c_n$ given observations $\tilde{c}_1, \ldots, \tilde{c}_n$ is the Kalman filter:
\begin{equation}
    \hat{c}_n = \hat{c}_{n|n-1} + K_n \left( \tilde{c}_n - \hat{c}_{n|n-1} \right), \label{eq:kalman_update}
\end{equation}
where $\hat{c}_{n|n-1} = \hat{c}_{n-1}$ from the random-walk dynamics and $K_n \in [0, 1]$ is the Kalman gain. Substituting gives:
\begin{equation}
    \hat{c}_n = (1 - K_n)\, \hat{c}_{n-1} + K_n\, \tilde{c}_n. \label{eq:kalman_sub}
\end{equation}
The Kalman gain evolves according to the Riccati recursion:
\begin{align}
    K_n &= \frac{P_{n|n-1}}{P_{n|n-1} + \sigma_v^2}, \label{eq:gain} \\
    P_{n|n-1} &= P_{n-1} + \sigma_w^2, \label{eq:predict} \\
    P_n &= (1 - K_n)\, P_{n|n-1}, \label{eq:update}
\end{align}
where $P_n = \mathrm{Var}(c_n - \hat{c}_n)$ is the estimation error variance. At steady state, $P_n \to P^*$ and $K_n \to K^*$, satisfying:
\begin{equation}
    K^* = \frac{P^* + \sigma_w^2}{P^* + \sigma_w^2 + \sigma_v^2}. \label{eq:ss_gain}
\end{equation}
Setting $\alpha = 1 - K^*$, the steady-state filter reduces to:
\begin{equation}
    \hat{c}_n = \alpha\, \hat{c}_{n-1} + (1 - \alpha)\, \tilde{c}_n, \quad \alpha \in (0, 1), \label{eq:ema}
\end{equation}
which is exactly exponential smoothing with factor $\alpha$. This establishes EMA as the optimal causal aggregation strategy under the linear-Gaussian assumption.

% The smoothing factor $\alpha$ encodes the noise-to-signal ratio: when observation noise dominates ($\sigma_v^2 \gg \sigma_w^2$), $\alpha \to 1$, placing more weight on the running estimate; when the state varies rapidly ($\sigma_w^2 \gg \sigma_v^2$), $\alpha \to 0$, trusting each new observation. In practice, $\alpha$ is treated as a hyperparameter rather than derived from estimated noise variances.

% The ablation in Table~\ref{tab:ablation_alpha} shows low sensitivity to $\alpha$, consistent with the limited frame-level variation observed in the projected coefficients after the sigmoid nonlinearity. We set $\alpha = 1$ in all experiments, which selects the initial frame's coefficients. Since wav2vec 2.0's Transformer contextualizes over the full utterance, even the first frame encodes global phonetic information, making this an effective operating point.

\begin{table}[t]
\centering
\caption{Ablation on FiLM conditioning location (VB-DEMAND). We ignore the 128 channel configuration for these experiments due to limited computational resources. Both the variants below are trained for 100 epochs.}
\label{tab:ablation_film}
\begin{tabular}{llccc}
\toprule
\textbf{Model} & \textbf{FiLM Location} & \textbf{PESQ\,$\uparrow$} & \textbf{STOI\,$\uparrow$} & \textbf{SI-SDR\,$\uparrow$} \\
\midrule
% \textsc{Ours-128}          & BN only   & x.xx & x.xx & x.xx \\
%\textsc{Ours-128 (Enc+BN)} & Enc + BN  & x.xx & x.xx & x.xx \\
% \midrule
\textsc{Ours-32}           & BN only   & \textbf{2.8634} & \textbf{0.8583} & \textbf{17.8051} \\
\textsc{Ours-32}  & Enc + BN  & 2.7941 & 0.8575 & 17.5364 \\
\textsc{Ours-32}  & Dec + BN  & 2.5048 & 0.8532 & 15.6289 \\
\bottomrule
\end{tabular}
\end{table}

\section{Experiments}

\subsection{Datasets and Setup}
\label{sec:Dataset_and_setup}
\textbf{Dataset.} We first train and evaluate on the VoiceBank-DEMAND (VB-DEMAND) dataset~\cite{valentini2017}, a widely used benchmark for speech enhancement. The training set consists of 11,572 utterances from 28 speakers in the VoiceBank corpus~\cite{veaux2013} mixed with 10 noise types from the DEMAND database~\cite{thiemann2013} at SNRs of 0, 5, 10, and 15\,dB. The test set contains 824 utterances from 2 held-out speakers mixed with 5 unseen noise types at SNRs of 2.5, 7.5, 12.5, and 17.5\,dB. All audio is sampled at 16\,kHz.

\textbf{Model configurations.} We evaluate two configurations of our proposed model: a full model with 128 channels (\textsc{Ours-128}) and a lightweight variant with 32 channels (\textsc{Ours-32}) in the U-Net. Both use a frozen wav2vec~2.0 base model~\cite{baevski2020} for feature extraction and a three-layer MLP as the FiLM generator. The diffusion process follows the  StoRM formulation~\cite{lemercier2023storm} with 30 reverse sampling steps. All models are trained for 100 epochs using Adam with a learning rate of 1e-4.

\textbf{Baselines.} We compare against four models: CDiffuSE~\cite{lu2022cdiffuse}, UNIVERSE \cite{serra2022}, UNIVERSE++ \cite{scheibler2024} and StoRM~\cite{lemercier2023storm} in both 128- and 32-channel configurations. For StoRM, we use the same 128/32-channel U-Net configurations as our model to enable a direct comparison of the effect of wav2vec~2.0 conditioning.

\textbf{Metrics.} We report PESQ (wideband)~\cite{rix2001pesq}, STOI~\cite{taal2011stoi}, and SI-SDR~\cite{leroux2019sdr} as intrusive metrics, and the DNSMOS components SIG, BAK, and OVRL~\cite{reddy2021dnsmos} as non-intrusive perceptual metrics. Intrusive metrics are the ones that require a ground truth waveform as reference for calculating scores whereas the non intrusive metrics don't.

\subsection{Ablation: Smoothing Coefficient $\alpha$}
\label{sec:ablation_alpha}

We investigate the sensitivity of enhancement quality to the hyperparameter $\alpha$. Table~\ref{tab:ablation_alpha} reports the average PESQ score on the VB-DEMAND test set, whose constituent files were split by SNR, for the 128-channel wav2vec 2.0 conditioned model across a range of $\alpha$ values. We observe the higher degree of smoothing (corresponding to higher value of $\alpha$) generally leads to better enhancement quality. Further, from Table \ref{tab:seen_intrusive} and Table \ref{tab:seen_dnsmos} we observe that exponential averaging achieves the right balance between intrusive and non intrusive metrics compared to mean pooling. This prompts us to use $\alpha = 1$ in all experiments in the subsequent sections.

\begin{table*}[t]
\centering
\caption{Speech enhancement results on the LibriMix dataset for the 32 channel configuration of the baseline and proposed models.}
\label{tab:librimix_results}
\begin{tabular}{llccccc}
\toprule
\textbf{Model} & \textbf{PESQ\,$\uparrow$} & \textbf{STOI\,$\uparrow$} & \textbf{SI-SDR\,$\uparrow$} & \textbf{SIG$\uparrow$} & \textbf{BAK\,$\uparrow$} & \textbf{OVRL\,$\uparrow$} \\
\midrule
\textsc{Ours-32}     & \textbf{2.0099} & \textbf{0.7836} & \textbf{11.9086} & $\mathbf{3.679 \pm 0.290}$ & $\mathbf{3.936 \pm 0.339}$ & $\mathbf{3.367 \pm 0.329}$\\
\textsc{StoRM-32}   & 1.6385 & 0.7409 & 9.6058 & $3.570 \pm 0.305$ & $3.046 \pm 0.537$ & $2.910 \pm 0.387$ \\
\bottomrule
\end{tabular}
\end{table*}

% NASE~\cite{hu2024nase},
\subsection{Results on VB-DEMAND}

Tables~\ref{tab:seen_intrusive} and~\ref{tab:seen_dnsmos} present results on the VB-DEMAND test set. As observed, both configurations of our model achieve strong performance across all metrics. Compared to StoRM at the same channel width, the addition of wav2vec~2.0 FiLM conditioning yields consistent improvements in PESQ and STOI for both configurations, confirming the value of the SSL-derived conditioning signal. Specifically, for the 128 channel variant we improve the PESQ score by 0.4 compared to StoRM. We observe through spectrogram analysis, that the degradation in SI-SDR scores is caused by aggressive noise suppression tendency of the proposed model.

% \begin{table*}[t]
% \centering
% \caption{Speech enhancement results on the LibriMix dataset for the 32 channel configuration of the baseline and proposed models.}
% \label{tab:librimix_results}
% \begin{tabular}{llccccc}
% \toprule
% \textbf{Model} & \textbf{PESQ\,$\uparrow$} & \textbf{STOI\,$\uparrow$} & \textbf{SI-SDR\,$\uparrow$} & \textbf{SIG$\uparrow$} & \textbf{BAK\,$\uparrow$} & \textbf{OVRL\,$\uparrow$} \\
% \midrule
% \textsc{Ours-32}     & \textbf{2.0099} & \textbf{0.7836} & \textbf{11.9086} & $\mathbf{3.679 \pm 0.290}$ & $\mathbf{3.936 \pm 0.339}$ & $\mathbf{3.367 \pm 0.329}$\\
% \textsc{StoRM-32}   & 1.6385 & 0.7409 & 9.6058 & $3.570 \pm 0.305$ & $3.046 \pm 0.537$ & $2.910 \pm 0.387$ \\
% \bottomrule
% \end{tabular}
% \end{table*}

% \subsection{Ablation: Smoothing Coefficient $\alpha$}
% \label{sec:ablation_alpha}

% We investigate the sensitivity of enhancement quality to the exponential smoothing coefficient $\alpha$. Table~\ref{tab:ablation_alpha} reports the average PESQ score on the VB-DEMAND test set, whose constituent files were split by SNR, for the 128-channel wav2vec 2.0 conditioned model across a range of $\alpha$ values. We observe the higher degree of smoothing (corresponding to higher value of $\alpha$) generally leads to better enhancement quality.

\subsection{Results on LibriMix}
% To further validate the generalizability of our proposed approach, we evaluate on the LibriMix dataset as a second, independently constructed dataset. This is important because it allows us to verify that our design choices are not overfitted to the acoustic conditions or noise characteristics of a single benchmark. 

LibriMix is derived from LibriSpeech and WHAM! noise sources, and we use the single-speaker configuration (Libri1Mix) with $min$ mixing mode at a 16 kHz sample rate, where clean utterances are mixed with realistic noise at various signal-to-noise ratios. The training set comprises 100 hours of noisy audio data spread over 13,900 waveforms, while the test set comprises 3000 utterances. The learning rate was maintained at 1e-4. We train the 32 channel variant of both our model and the baseline for 87 epochs to enable a fair comparison. We chose this number because the baseline model converged after these many epochs with a patience of 50 epochs. Results on both intrusive and non intrusive metrics are presented in Table \ref{tab:librimix_results}.

\subsection{Ablation: FiLM Conditioning Location}
Table~\ref{tab:ablation_film} compares bottleneck-only conditioning with a variant (\textsc{Ours-Enc+BN}) that applies FiLM at all four encoder resolutions as well. Extending conditioning beyond the bottleneck degrades performance, which we attribute to a scale mismatch: early encoder layers process fine-grained spectro-temporal details that are disrupted by high-level semantic modulation, whereas the bottleneck operates at an abstraction level naturally aligned with wav2vec~2.0 features.

\subsection{Computational Cost and Inference Time}

Our observations reveal that the U-Net dominates the total computational cost: for the 128-channel configuration, a single U-Net forward pass requires 312\,GFLOPs, which is multiplied by $N$\,=\,30 reverse steps, yielding 9360\,GFLOPs for the iterative sampling alone. For the 32-channel configuration this number is 19.5 GFLOPs and 592 GFLOPs respectively. By contrast, the wav2vec~2.0 base requires approximately 13.27\,GFLOPs for a 1-second utterance, which is a one-time cost that amounts to only 0.1\% of the total inference budget for \textsc{Ours-32}. The FiLM generator contributes only 0.0058\,GFLOPs/sec, which is negligible by comparison. Further, Table~\ref{tab:inference} reports real time fator (RTF) on the VB-DEMAND test set, measured on a single NVIDIA GTX 3090 GPU. The total duration of the test set is roughly 35 minutes. In particular, \textsc{Ours-128} incurs only a 10.4\% increase in RTF over StoRM-128, and \textsc{Ours-32} adds 52.7\% over StoRM-32 yet \textsc{Ours-32} achieves an RTF of 0.55, demonstrating that this configuration is capable of real time capabilities.

% Table~\ref{tab:inference} reports inference times on the VB-DEMAND test set, measured on a single NVIDIA GTX 3090 GPU. Despite incorporating a frozen wav2vec~2.0 encoder ($\sim$95M parameters), the additional cost is modest: the wav2vec~2.0 forward pass is performed only once per utterance (13.27 GFLOPs per second of audio), whereas the iterative reverse diffusion accounts for the vast majority of computation (e.g., 592 GFLOPs for 30 U-Net passes in the 32-channel model). The per-step U-Net cost is identical between StoRM and our model at the same channel width, as FiLM adds only element-wise operations at the bottleneck. \textsc{Ours-32} achieves an RTF of 0.55, confirming real-time capability.

\subsection{Spectrogram Analysis}

\begin{figure}[htbp]
  \centering

  \begin{subfigure}{0.5\textwidth}
    \centering
    \includegraphics[width=\linewidth]{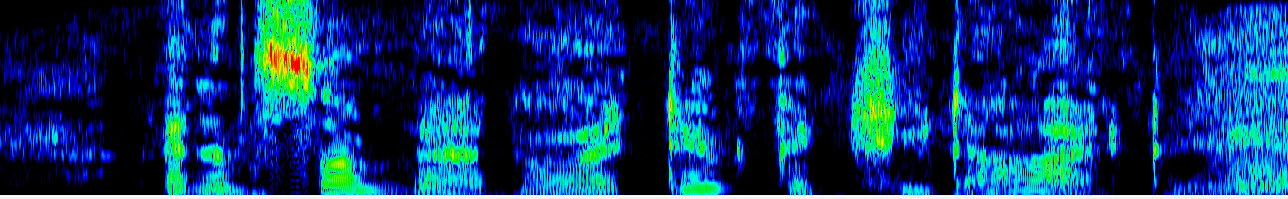}
    \caption{}
    \label{fig:sub1}
  \end{subfigure}

  \begin{subfigure}{0.5\textwidth}
    \centering
    \includegraphics[width=\linewidth]{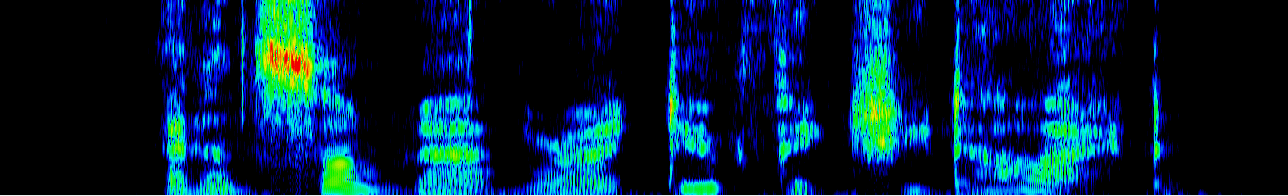}
    \caption{}
    \label{fig:sub2}
  \end{subfigure}

  \begin{subfigure}{0.5\textwidth}
    \centering
    \includegraphics[width=\linewidth]{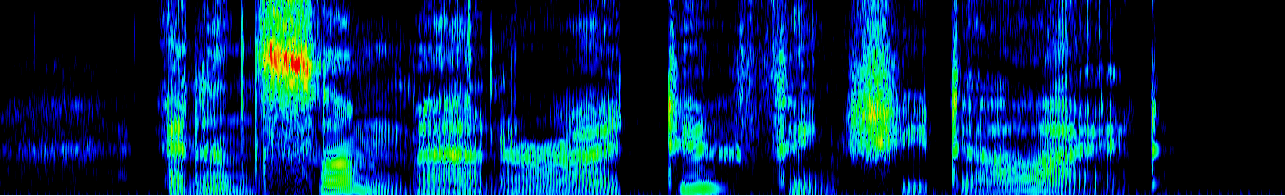}
    \caption{}
    \label{fig:sub3}
  \end{subfigure}

  \caption{Sample spectrograms of a LibriMix test set utterance enhanced by (a) baseline StoRM-128, (b) Ours-128 and (c) the ground truth clean waveform.}
  \label{fig:overall_spectrograms}
\end{figure}

We present sample output spectrograms in Figure \ref{fig:overall_spectrograms} to corroborate the objective metrics and revealing distinct qualitative advantages of the proposed film conditioned architecture over the baseline model. The ground truth spectrogram (Figure \ref{fig:overall_spectrograms}(c)) establishes a baseline of clean acoustic features, characterized by sharp formant transitions and distinct harmonic structures during voiced segments. The baseline model (Figure \ref{fig:overall_spectrograms}(a)), while successful at basic denoising, suffers from noticeable spectral blurring and over-smoothing. This is particularly evident in the mid-to-high frequency bands, where fine harmonic details are smeared and high-frequency fricatives are heavily attenuated, often leading to a poor perceptual quality. In contrast, the proposed film conditioned model (Figure \ref{fig:overall_spectrograms}(b)) demonstrates superior spectral fidelity. By dynamically modulating the feature maps, the proposed model successfully preserves the sharp contours of the formants and restores the fine harmonic structure lost by the baseline. Furthermore, the boundaries between speech and non-speech regions are distinctly crisper, indicating that FiLM conditioning enables the network to more selectively suppress noise without aggressively compromising the underlying speech components. This visually confirms that the proposed approach effectively bridges the existing gap with ground-truth signal reconstruction.

\section{Conclusion}
We presented a method for conditioning diffusion-based speech enhancement on wav2vec 2.0 representations via FiLM modulation at the U-Net bottleneck, with temporally smoothed coefficients. Evaluation on VoiceBank-DEMAND and LibriMix benchmarks show  competitive performance on intrusive and non intrusive metrics against unconditioned baselines with negligible computational cost overhead, and ablations confirm that bottleneck-only conditioning is most effective. Future work includes evaluation on larger benchmarks, formal subjective listening tests and experimenting with WavLM/ HuBERT feature extractors.

\section{Generative AI Use Disclosure}
Generative AI tools were used to refine and polish portions of the manuscript text. They were not used for idea formulation, experimental design, or generating significant portions of the text from scratch.

\bibliographystyle{IEEEtran}
\bibliography{mybib}

\end{document}